# Two-Phase Emission Detector for Measuring Coherent Neutrino-Nucleus Scattering

Chris Hagmann and Adam Bernstein

*Abstract*—Coherent scattering is a flavor-blind, high-rate, as yet undetected neutrino interaction predicted by the Standard Model. We propose to use a compact (kg-scale), two-phase (liquid–gas) argon ionization detector to measure coherent neutrino scattering off nuclei. In our approach, neutrino-induced nuclear recoils in the liquid produce a weak ionization signal, which is transported into a gas under the influence of an electric field, amplified via electroluminescence, and detected by phototubes or avalanche diodes. This paper describes the features of the detector, and estimates signal and background rates for a reactor neutrino source. Relatively compact detectors of this type, capable of detecting coherent scattering, offer a new approach to flavor-blind detection of man-made and astronomical neutrinos, and may allow development of compact neutrino detectors capable of nonintrusive real-time monitoring of fissile material in reactors.

## I. INTRODUCTION

COHERENT neutrino-nucleus scattering is a famous but as yet untested prediction of the Standard Model [1]. The process is mediated by neutral currents (NC), and hence is flavor-blind. Despite having relatively high rates, neutrino-nucleus scattering is difficult to observe because its only signature is a small nuclear recoil of energy $\sim$ keV (for MeV neutrinos), requiring a low detector threshold. Over the past two decades, a number of authors have suggested low-temperature calorimeters [1], [2], gas detectors [3], and germanium ionization detectors [4] for measuring neutrino-nucleus scattering. In this paper, we study a two-phase (gas-liquid) ionization detector, which combines low energy threshold with large event rates.

Coherent neutrino-nucleus scattering has the cross section [1] $\sigma \sim 0.4 \times 10^{-44} N^2 (E_\nu/\text{MeV})^2 \text{cm}^2$, where $N$ is the neutron number, and $E_\nu$ is the neutrino energy. This formula is valid for neutrino energies up to about 50 MeV, and thus applies to reactor, solar, and supernova neutrinos. For a fixed neutrino energy, the recoil spectrum falls linearly. The average energy is

$$\langle E_r \rangle = \frac{1}{3} E_r^{\max} = 716 \text{ eV} \left( \frac{\left(\frac{E_\nu}{\text{MeV}}\right)^2}{A} \right) \quad (1)$$

where $A$ is the atomic number of the target nucleus.

Manuscript received December 3, 2003; revised March 11, 2004 and May 7, 2004. This work was supported in part by the U.S. Department of Energy by University of California, Lawrence Livermore National Laboratory, under Contract W-7405-ENG-48.

The authors are with the Lawrence Livermore National Laboratory, Livermore, CA 94550 USA (e-mail: hagmann1@llnl.gov, bernstein3@llnl.gov).

Digital Object Identifier 10.1109/TNS.2004.836061

It is well known however, that recoiling atoms are less effective in producing primary ionization or scintillation than electrons of the same energy. The ratio of the ionization (and/or scintillation) yield from atomic projectiles to that from electrons, referred to as the quench factor $q$, generally decreases with energy and is material dependent. For example, measured $q$ factors in silicon [5] decrease from 0.41 to 0.26, for recoil energies of 21 keV and 3.3 keV, respectively. An even smaller quench factor of $q = 0.15$ was reported for germanium [6], at a recoil energy $E_r = 254$ eV.

A signal consisting of only a few electrons or photons is below threshold for conventional solid or liquid state detectors without internal amplification. Hence we propose a two-phase (gas-liquid) argon emission detector with an electroluminescence gap in the gas to provide gain. This scheme combines a large target density in the liquid with the capability of sensing single electrons. Its moderate cost and scalability, as compared to calorimetric detectors, make this technology a promising approach to NC based detection of reactor and astronomical neutrinos.

## II. RECOIL RATE AND IONIZATION YIELD

An attractive attribute of neutrino coherent scattering is its relatively large cross section compared to inverse beta decay. For reactor neutrinos [$\Phi \sim 6 \times 10^{12}$ cm$^{-2}$s$^{-1}$ at $\sim 25$ m from a 3-gigawatt thermal (GWt) core], the expected event rates before detection efficiencies are 56 kg$^{-1}$day$^{-1}$ for coherent scattering off argon, compared to 2.8 kg$^{-1}$day$^{-1}$ for the inverse beta decay reaction in $(\text{CH})_n$. Here we assumed a typical fuel mix of 61.9% $^{235}$U, 6.7% $^{238}$U, 27.2% $^{239}$Pu, and 4.2% $^{241}$Pu, with neutrino spectra and mix parameters taken from [7], [8]. Fig. 1 shows the expected argon recoil spectra, obtained by convoluting the reactor neutrino spectrum [7] with the theoretically predicted nuclear recoil energy distribution [1]. Although the average recoil energy in argon is $\sim 200$ eV, the majority of the recoil events do not produce primary ionization or excitation because of quenching.

In order to estimate the amount of ionization produced by recoils, we performed a Monte Carlo simulation of the atomic collision cascade. Our computer calculations are based on the transport of ions in matter (TRIM) code [9], which models the collisions as a series of binary events separated by a path length $L = n^{-1/3}$ (= $3.6 \times 10^{-8}$ cm in liquid argon), where $n$ is the atomic number density. For each collision step, the impact parameter is sampled by randomly choosing a point within a disk of area $\sigma_{\text{geo}} = n^{-2/3}$. The scattering angle and hence the elastic energy transfer is determined by a Molière inter-atomic potential. Inelastic interactions were modeled by sampling the measured Ar–Ar ionization and excitation cross sections for each collision with probabilities $p_{\text{ion}} = \sigma_{\text{ion}}/\sigma_{\text{geo}}$ and $p_{\text{exc}} = \sigma_{\text{exc}}/\sigma_{\text{geo}}$. Fig. 2 depicts the inelastic argon cross sections compiled by Phelps [10]. The inelastic energy losses were accounted for in the energy budget of each cascade as follows. For collision energies < 1 keV, ionization is primarily produced via the creation of the auto-ionizing state $(1s^2 2s^2 2p^6 3s^1 3p^6 4s^1)$ with excitation energy $\sim 25$ eV, leading to the subsequent emission of an Auger electron of energy $\sim 9.4$ eV [11]. The exiting Ar$^+$ projectile neutralizes quickly by charge exchange ($\sigma_{\text{cx}} \sim 10^{-15}$ cm$^2$), with its energy reduced by the ionization potential of liquid argon ($I_p = 14.3$ eV [12]). The inelastic excitation energy loss ($\Delta E \sim 25$ eV) is shared evenly between the colliding bodies, while the energy loss due to charge exchange ($\Delta E = I_p$) is randomly assigned to one of the outgoing projectiles. Excited argon atoms (Ar$^*$) are assumed to be created in state $(1s^2 2s^2 2p^6 3s^2 3p^5 4s^1)$ with energy $\Delta E \sim 12$ eV. The primary projectile and all energetic secondary particles produced in the cascade are followed till their energies drop below the inelastic reaction threshold ($\sim 25$ eV in the laboratory frame).

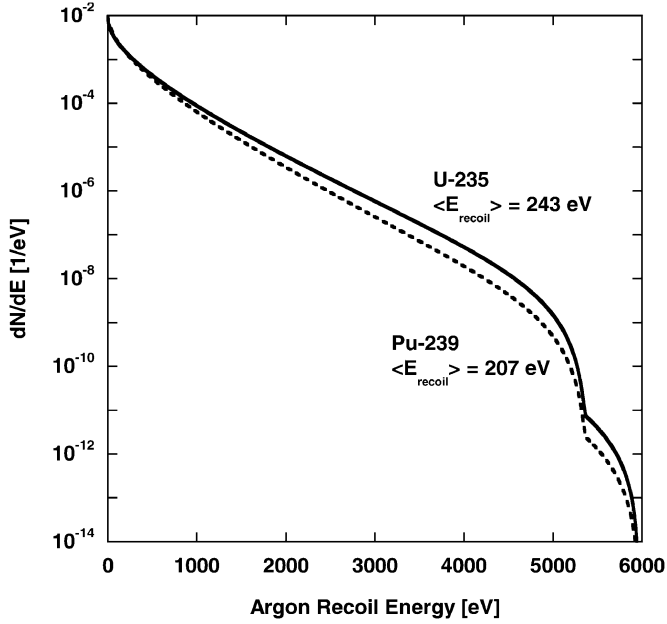

Fig. 1. Nuclear recoil spectra predicted for $^{235}$U and $^{239}$Pu fission neutrinos scattering coherently off natural argon.

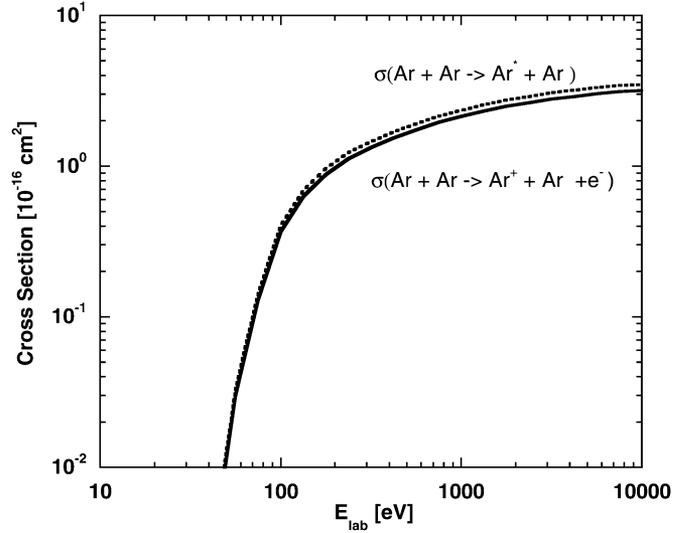

Fig. 2. Excitation and ionization cross sections for Ar + Ar collisions.

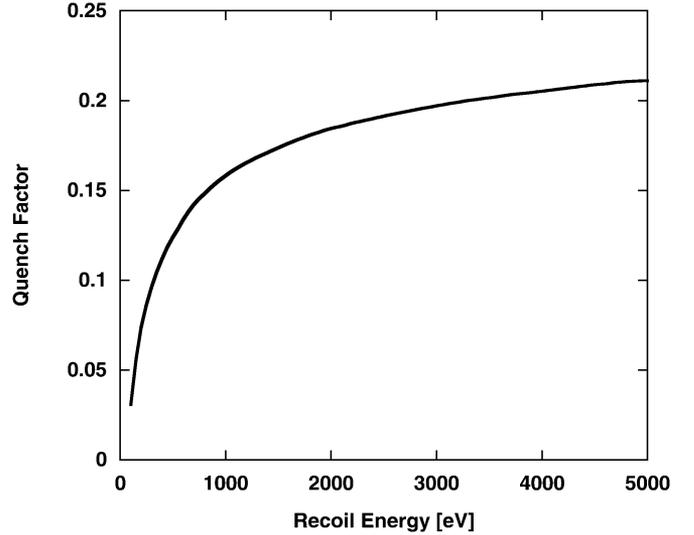

Fig. 3. Calculated quench factor of liquid argon, defined as the ratio of the nuclear to electron recoil induced inelastic (ionization or excitation) yield.

The Monte Carlo code allowed us to calculate the average ionization and excitation yields as a function of recoil energy $E$, and thus determine the Ar quench factor (shown in Fig. 3), defined here as

$$q(E) \equiv \frac{N_{\text{ion}}^{\text{nucl}}(E) + N_{\text{exc}}^{\text{nucl}}(E)}{N_{\text{ion}}^{\text{elec}}(E) + N_{\text{exc}}^{\text{elec}}(E)} \quad (2)$$

where $N_{\text{ion}}$ and $N_{\text{exc}}$ are the energy-dependent average ionization and excitation numbers. Kubota et al. [13] measured the yields from electron recoils

$$N_{\text{ion}}^{\text{elec}}(E) + N_{\text{exc}}^{\text{elec}}(E) \approx 1.21 \frac{E}{W} \quad (3)$$

and related them to $W$, the average electronic energy required to produce an electron-ion pair. In liquid argon, $W$ has a value of $\approx 23$ eV [14], [15].

We obtained the reactor neutrino ionization spectrum, depicted in Fig. 4, by convolving the ionization efficiency with the recoil spectrum. About 29% of all recoils produce at least a single electron-ion pair. In addition, a similar number of Ar$^*$ excitons are created with an identical number spectrum. Some of the excitation can be converted into ionization via doping with xenon. Since the Ar$^*$ exciton energy exceeds the ionization potential of xenon in liquid argon ($I_p(\text{Xe}) \sim 10.6$ eV), the secondary ionization process (Ar$^*$ + Xe $\rightarrow$ Ar + Xe$^+$ + $e^-$ is energetically allowed. Experimentally [13], the probability for this Penning mechanism to occur is

$$p_{\text{Penning}} = \frac{1.44 f_{\text{Xe}}[\%]}{1 + 1.44 f_{\text{Xe}}[\%]} \quad (4)$$

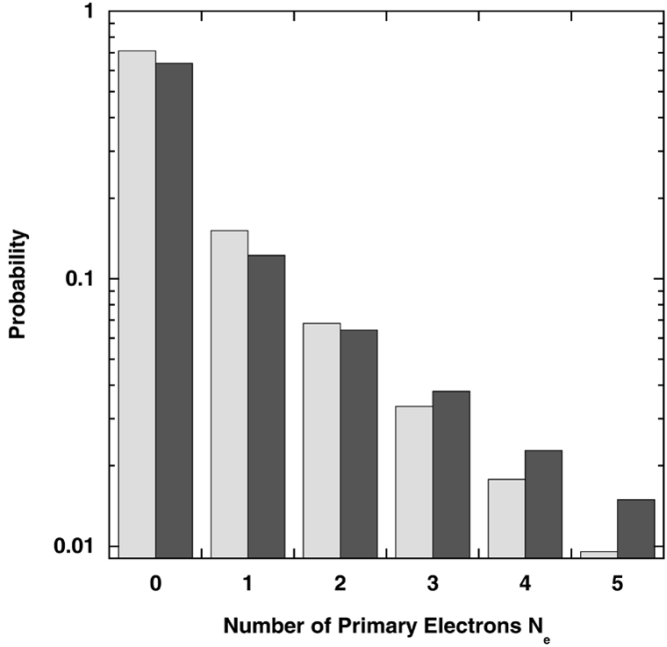

Fig. 4. Ionization number spectrum arising from reactor neutrinos. The light-shaded bins represent the spectrum for pure argon, the dark-shaded bins are for argon doped with $\sim 1\%$ xenon.

TABLE I
SIGNAL PARAMETERS FOR A REACTOR NEUTRINO FLUX OF $6 \times 10^{12}$ cm$^{-2}$s$^{-1}$ AT A DISTANCE OF 25 M FROM A TYPICAL 3-GWt CORE

| | |
|---|---|
| Average argon recoil energy | 234 eV |
| Number of recoils per day per 10 kg of argon | 560 |
| Fraction of ionizing recoils in pure argon | 29 % |
| Fraction of ionizing recoils with 1% xenon doping | 36 % |

where $f_{Xe}$ is the xenon concentration in liquid argon. The number of free electrons created in argon is consequently enhanced to

$$N_e = N_{ion} + p_{Penning} N_{exc}. \qquad (5)$$

Table I summarizes the expected recoil and ionization rates.

### III. DETECTOR SCHEME

Our proposed detector is similar to those currently being developed for WIMP dark matter experiments [16]–[20]. Emission detectors house two phases (liquid-gas or liquid-solid) of a noble element in a single cell [21]. They may combine a large detector mass with a low detection threshold, and are ideally suited for measuring rare events in the kiloelectronvolt range. The primary ionization event most likely takes place in the condensed phase of the detector, where free electrons are produced. An applied electric field causes the electrons to drift toward the phase boundary and cross into the gas, where the charge signal is converted to an ultraviolet light signal via proportional scintillation. Geminate recombination and capture on electronegative impurities, such as $O_2$, may lead to electron loss. The rate of the former is proportional to the product of the positive and negative charge densities, and thus small for weak ionization events. The latter process can be made negligible by keeping the transfer time smaller than the free electron lifetime. Bakale et al. [22] measured an attachment lifetime of

$$\tau_a \approx \frac{10^{-11}}{\frac{[O_2]}{[Ar]}} s \qquad (6)$$

and lifetimes of a few ms are routinely achieved using commercially available purification systems.

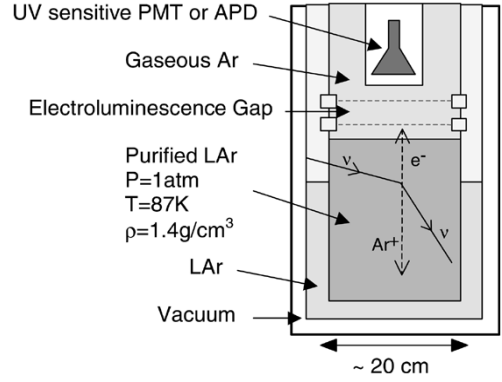

Fig. 5. Schematic of proposed detector.

The electron transfer time (between primary event time till detection in the gas region) is usually dominated by the drift time $\tau_d$ (in the liquid), and the phase boundary crossing time $\tau_x$. The electron drift velocity in liquid argon is electric field dependent. For the range $10^2$ V/cm $< \varepsilon < 10^3$ V/cm, the drift time $\tau_d$ over a distance $L_d$ (in cm) is approximately given by [12]

$$\tau_d \approx 10^{-5} L_d \left( \frac{\varepsilon}{300 \frac{V}{cm}} \right)^{-0.64} s. \qquad (7)$$

The electronic potential barrier height of the liquid-gas interface in argon is $\sim 0.2$ V, and the electrons are transferred into the gas by field-assisted thermionic emission. Borghesani et al. [21] determined a crossing time of

$$\tau_x \approx \frac{0.1}{\varepsilon} e^{-0.06 \varepsilon^{1/2}} s \qquad (8)$$

where $\varepsilon$ has units of V/cm.

Once in the gas phase, the electrons traverse an electroluminescence gap defined by two parallel grids with an applied potential of a few kilovolts. Inelastic collisions create Ar$_2^*$ molecules which decay radiatively, emitting UV photons of energy $\sim 10$ eV. Both singlet and triplet states are created, with lifetimes of 4 ns and 3 ìs respectively [24]. Dias et al. [25] have extensively modeled the scintillation efficiency as a function of the reduced field $\varepsilon/n$. The light conversion efficiency, i.e., the fraction of electric potential energy converted into scintillation energy, rises from the threshold value $(\varepsilon/n)_c = 3 \times 10^{-17}$ Vcm$^2$ roughly linearly to $\sim 50\%$ at $(\varepsilon/n) = 7 \times 10^{-17}$ Vcm$^2$. Gain values of a few hundred photons per electron with a cm scale gap are typical. Lastly the UV light needs to be collected with high efficiency to enable detection of single electrons. Both large-area, UV-sensitive phototubes and windowless avalanche diodes are attractive options. Fig. 5 shows a schematic of the detector we envisage for this experiment.

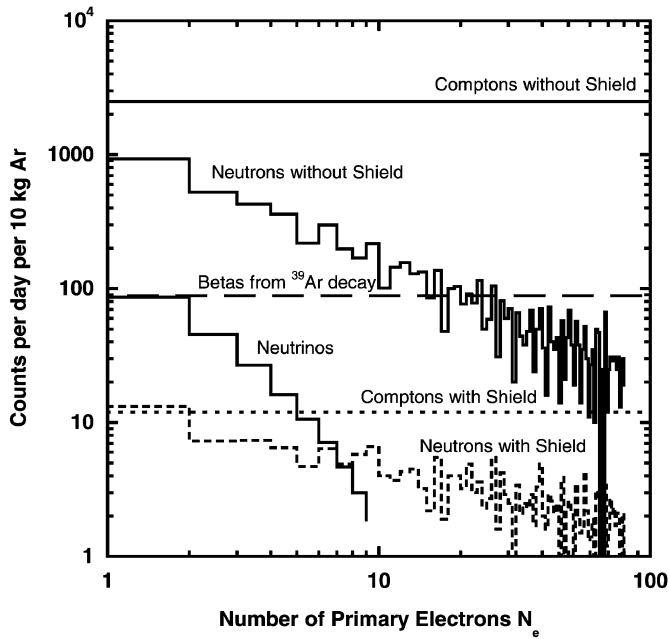

Fig. 6. Monte Carlo simulation of gamma and neutron induced detector backgrounds. Characteristic gammas ($E > 50$ keV) were sourced isotropically within a 20 cm thick concrete wall surrounding the argon detector. The concrete had U/Th/K activities of 104/47/533 mBq/cm$^3$ respectively. With regard to neutrons, an isotropic source with flux $\sim 10^5$ m$^{-2}$s$^{-1}$ (detector site depth = 20 mwe), and a $1/E$ spectrum from thermal energies up to 20 MeV was used in the simulation. A reduction of the external backgrounds by $\sim 100$ is achieved with a layered shield of 2 cm of lead (inner) and 10 cm of borated polyethylene (outer). A reactor neutrino flux of $6 \times 10^{12}$ cm$^{-2}$s$^{-1}$ is assumed.

## IV. CALIBRATION

The detector could be calibrated for example by fast neutron elastic scattering [5] or by thermal neutron capture [3], [6]. The latter is well suited for producing sub-keV recoil energies in argon. The $^{40}$Ar $(n,\gamma)^{41}$Ar reaction has a $Q = 6.098$ MeV, which is released into two characteristic gamma rays ($E_1 = 5.582$ MeV, $E_2 = 0.516$ MeV) with a branching ratio of $\sim 10\%$. The first gamma gives a recoil $E_r = 1/2\, E_1^2/M_{Ar} = 415$ eV, while the second one could be used for tagging the recoil.

## V. BACKGROUNDS

Low energy (few primary electrons) events may be caused by (1) small angle Compton recoils from internal and external gamma radioactivity, (2) internal low-energy beta activity, and (3) nuclear recoils arising from neutron scattering or capture. The neutrons may come from ambient radioactivity present in nearby material, or from muon interactions with surrounding rock and detector materials. The cosmic muons themselves have energy deposits $\gg 1$ keV and can be vetoed. Backgrounds due to external gamma/neutron activity can be readily reduced by lead/polyethylene shielding and operating the detector at a shallow underground site near the reactor. The strongest internal radioactivity is from the beta decay of $^{39}$Ar ($\tau_{1/2} = 269\, y$, end point $E = 565$ keV). The measured $^{39}$Ar activity is $\sim 1$ Bq/kg in natural argon [26], [27]. The Coulomb corrected electron spectrum, $dN/dE = F(Z,E)\Phi(E)$, is approximately constant for low electron energies [28]. Here $F(Z,E)$ is the Fermi function, $Z$ is the charge number of the daughter nucleus, and $\Phi(E)$ is the statistical factor of beta decay. Using this expression, we estimate a differential beta activity of $\sim 5$ mBq/(keV $\cdot$ kg) at the low energy end. Fig. 6 shows the estimated background rates in a bare detector and for a dual-shield configuration.

## ACKNOWLEDGMENT

The authors would like to thank A. Bolozdynya for illuminating discussions.

## REFERENCES

[1] A. Drukier and L. Stodolsky, "Principles and applications of a neutral-current detector for neutrino physics and astronomy," *Phys. Rev. D, Part. Fields*, vol. D30, pp. 2295–2309, 1984.

[2] B. Cabrera, L. M. Krauss, and F. Wilczek, "Bolometric detection of neutrinos," *Phys. Rev. Lett.*, vol. 55, pp. 25–28, 1985.

[3] P. S. Barbeau, J. I. Collar, J. Miyamoto, and I. Shipsey, "Toward coherent neutrino detection using low-background micropattern gas detector," *IEEE Trans. Nucl. Sci.*, vol. 50, pp. 1285–1289, Oct. 2003.

[4] A. S. Starostin and A. G. Beda, "Germanium detector with internal amplification for investigating rare processes," *Phys. Atom. Nucl.*, vol. 63, pp. 1297–1300, 2000.

[5] G. Gerbier et al., "Measurement of the ionization of slow silicon nuclei in silicon for the calibration of a silicon dark-matter detector," *Phys. Rev. D, Part. Fields*, vol. D42, pp. 3211–3214, 1990.

[6] K. W. Jones and H. W. Kraner, "Energy lost to ionization by 254-eV $^{73}$Ge atoms stopping in Ge," *Phys. Rev.*, vol. A11, pp. 1347–1353, 1975.

[7] P. Vogel and J. Engel, "Neutrino electromagnetic form factors," *Phys. Rev.*, vol. D39, pp. 3378–3383, 1989.

[8] G. Zacek et al., "Neutrino-oscillation experiments at the Gösgen nuclear power reactor," *Phys. Rev. D, Part. Fields*, vol. D34, pp. 2621–2636, 1986.

[9] J. P. Biersack and L. G. Haggmark, "A Monte Carlo computer program for the transport of energetic ions in amorphous targets," *Nucl. Instrum. Methods*, vol. 174, pp. 257–320, 1980.

[10] A. V. Phelps, "Cross sections and swarm coefficients for nitrogen ions and neutrals in N$_2$ and argon ions and neutrals in Ar for energies from 0.1 eV to 10 keV," *J. Phys. Chem. Ref. Data*, vol. 20, no. 3, pp. 557–573, 1991.

[11] G. Gerber, R. Morgenstern, and A. Niehaus, "Ionization processes in slow collisions of heavy particles II. Symmetrical systems of the rare gases He, Ne, Ar, Kr," *J. Phys. B, Atom. Molec. Phys.*, vol. 6, pp. 493–510, 1973.

[12] T. Doke, "Fundamental properties of liquid argon, krypton, and xenon as radiation detector material," *Portugal Phys.*, vol. 12, pp. 9–48, 1981.

[13] S. Kubota et al., "Evidence of the existence of exciton states in liquid argon and exciton-enhanced ionization from xenon doping," *Phys. Rev. B, Condens. Matter*, vol. B13, pp. 1649–1653, 1976.

[14] T. Takahashi, S. Konno, and T. Doke, "The average energies, $W$, required to form an ion pair in liquefied rare gases," *J. Phys. C, Solid State Phys.*, vol. 7, pp. 230–240, 1974.

[15] S. S.-S. Huang, N. Gee, and G. R. Freeman, "Ionization of liquid argon by x-rays: Effect of density on electron thermalization and free ion yields," *Radiat. Phys. Chem.*, vol. 37, pp. 417–421, 1991.

[16] D. Y. Akimov et al., "Development of a two-phase xenon dark matter detector," *Physics of Atom. Nuclei*, vol. 66, pp. 497–499, 2003.

[17] D. B. Cline, "A WIMP detector with two-phase liquid xenon," *Nucl. Phys. B, Proceedings Supplements*, vol. 87, pp. 114–116, 2000.

[18] E. Aprile et al., "XENON: A 1 tonne liquid xenon experiment for a sensitive dark matter search," presented at the Int. Workshop on Technique and Application of Xenon Detectors, Y. Suzuki, M. Nakahata, Y. Koshio, and S. Moriyama, Eds., Tokyo, Japan, Dec. 2001.

[19] M. Yamashita, T. Doke, J. Kikuchi, and S. Suzuki, "Double phase (liquid/gas) xenon scintillation detector for WIMP's direct search," *Astropart. Phys.*, vol. 20, pp. 79–84, 2003.

[20] C. Rubbia, "WARP liquid argon detector for dark matter," presented at the 6th UCLA Symp. Sources and Detection of Dark Matter and Dark Energy in the Universe, Marina del Rey, CA, Feb. 18–20, 2004.

[21] A. Bolozdynya, "Two-phase emission detectors and their applications," *Nucl. Instrum. Methods*, vol. A422, pp. 314–320, 1999.

[22] G. Bakale, U. Sowada, and W. F. Schmidt, "Effect of an electric field on electron attachment in $SF_6$, $N_2O$, $O_2$ in liquid argon and xenon," *J. Phys. Chem.*, vol. 80, pp. 2556–2559, 1976.
[23] A. F. Borghesani, G. Carugno, M. Cavenago, and E. Conti, "Electron transmission through the Ar liquid-vapor interface," *Phys. Lett.*, vol. A149, pp. 481–484, 1990.
[24] N. Schwentner, E.-E. Koch, and J. Jortner, *Electronic Excitations in Condensed Rare Gases*. Berlin, Germany: Springer-Verlag, 1985, pp. 129–129.
[25] T. H. V. T. Dias, A. D. Stauffer, and C. A. N. Conde, "A unidimensional Monte Carlo simulation of electron drift velocities and electroluminescence in argon, krypton, and xenon," *J. Phys. D, Appl. Phys.*, vol. 19, pp. 527–545, 1986.
[26] H. H. Loosli, "A dating method with $^{39}$Ar," *Earth Planet. Sci. Lett.*, vol. 63, pp. 51–62, 1983.
[27] W. Kutschera *et al.*, "Long-lived noble gas radionuclides," *Nucl. Instrum. Methods*, vol. B92, pp. 241–248, 1994.
[28] C. S. Wu and S. A. Moszkowski, *Beta Decay*. New York: Interscience, 1966, p. 30.